# Floating Point HUB Adder for RISC-V Sargantana Processor

Gerardo Bandera[1],[*], Javier Salamero[2], Miquel Moreto[2,3] and Julio Villalba[1]

[1]Computer Architecture Dept., University of Málaga
[2]BSC-CNS - Barcelona Supercomputing Center, [3]Computer Architecture Dept, UPC

**Abstract**

*HUB format is an emerging technique to improve the hardware and time requirement when round to nearest is needed. On the other hand, RISC-V is a open source ISA that an important number of companies are using in their designs currently. In this paper we present a tailored floating point HUB adder that has been implemented in the Sargantana RISC-V processor.*

## HUB format background

In this section we present a basic description of the HUB format, although the mathematical foundations and an in-depth analysis of the format can be found in [1]. Now we briefly summarize this new real number representation format defined in [1] and particularize it for the floating-point normalized numbers.

HUB format is based on shifting the numbers that can be exactly represented under conventional formats by adding a bias which equals half unit in the last-place (ULP). This shifting could be also interpreted as appending a hidden least significant bit set to one to the conventional number stream (which represents the bias). A floating-point HUB number is similar to a regular one but its significand follows the HUB format. Thus, the exponent and the sign is the same as convetional. Let us define $x$ as a floating-point HUB number, which is represented by the triple $(S_x, M_x, E_x)$ such that $x = (-1)^{S_x} M_x 2^{E_x}$, where the significand $M_x$ is a HUB magnitude. A normalized HUB significand fulfills that $1 < M_x < 2$. Thus, the normalized HUB significand $M_x$ is

$$M_x = 1 + \left[\sum_{i=1}^{f} M_{x_i} \cdot 2^{-i}\right] + 2^{-f-1} \qquad (1)$$

where $2^{-f-1}$ is the bias. In this expression we define the *representative* form of the normalized HUB significand as the set of $M_{x_i}$ in expression (1), that is $M_x = (M_{x_1}, M_{x_{-1}}, M_{x_{-2}}, \cdots, M_{x_{-f}})$ (composed by $f$ bits). Taking into account that both the MSB and the LSB are 1 (see expression (1)), we define the *operational* form of a normalized HUB significand as the following $f + 2$ bits:

$$M_x = 1.M_{x_{-1}} M_{x_{-2}} \cdots M_{x_{-f}} 1 \qquad (2)$$

The representative version is used for storage whereas the operational version is required to operate with HUB numbers. We can see that the least significant bit (LSB) of the operational form of a nonzero HUB number is always equal to 1, and it is implicit in the format (similar situation takes place for the most significant bit (MSB) of the significand in the IEEE normalized floating-point numbers). Let ILSB denote the implicit LSB of a the operational HUB significand.

Given a standard floating-point system with a normalized significand, the counterpart HUB floating point system has the same precision and accuracy [1].

The most outstanding feature of the HUB format is that round to nearest is performed by truncation. In the conventional format round to nearest is carried out by adding one to the position of the rounding bit of the final normalized result. Moreover, after this operation an overflow can also be produced, which involves a shift operation of one bit and an update of the exponent. Thus, a specific hardware module is used in conventional. In HUB, this module is not required any more.

## FP HUB adder for Sargantana

The processor used in this work is Sargantana [2], a 64 bit in-order Linux-capable RISC-V CPU that implements the RV64G ISA (see figure 1). For accelerating domain-specific applications, Sargantana uses a Single Instruction Multiple Data (SIMD) unit and supports the vector instructions defined in the vector extension RVV 0.7.1. In addition, it implements custom application specific instructions. The CPU has a 7-stage pipeline that implements register renaming, out-of-order write-back, and a non-blocking memory pipeline. It has two first level caches: an instruction cache of 16KB, and a non-blocking data cache of 32KB. The system also has a 512KB L2 cache outside the CPU.

Figure 2 shows the main modules required to carry out the floating point addition for conventional and for HUB. Since round to nearest operation is carried out by truncation in HUB, the result obtained after normalization module in figure 2 is the final result and not any other operation is required. Thus, the module Rounding (crossed out in the figure 2) is not required for the HUB implementation in Sargantana.

[*]Corresponding author: `mailto:gbandera@uma.es`

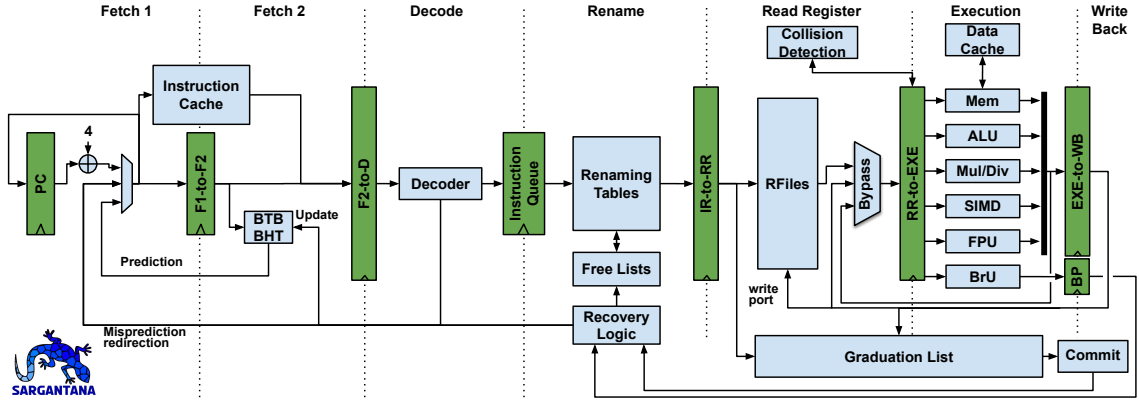

**Figure 1:** *Architecture of the Sargantana processor*

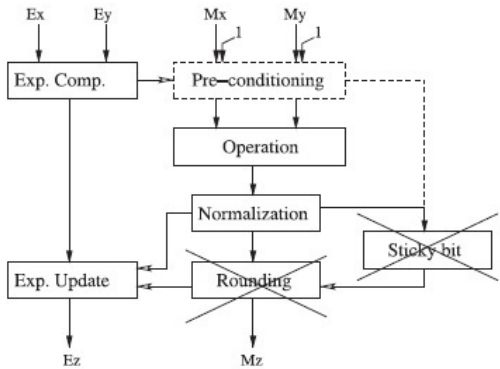

**Figure 2:** *Main modules in conventional and HUB FP adders (X-> prevented in HUB)*

For applications requiring conventional addition with round-to-nearest, three extra fractional bits are needed: a guard bit, a rounding bit and a sticky bit [3]. Since the round-to-nearest for HUB format is carried out by truncation, the rounding bit is not required anymore. Moreover, in conventional it is necessary calculate the sticky bit (this bit represents, in some way, the bits beyond the rounding bit and is needed for effective subtraction when the operands are not aligned). For HUB number, because we know that the LSB of the shifted operand is always 1 (that is the ILSB), the sticky bit is always 1 and it is not necessary a module to calculate it, as shown in figure 2.

Thus, in spite of having an extra bit in the operational form (the ILSB), this extra bit of a HUB number is compensated by the lack of a specific rounding bit.. Moreover, a guard bit is not necessary and the sticky bit, when required, is always 1. As consequence, for applications where round to nearest is required, the data path of the HUB version has one bit less than that its conventional counterpart

Our starting point is the FPU adder of the RISC-V Sargantana processor which has been modified to meet the HUB format specifications. Unlike the 6-stage original FPU adder, our floating point HUB adder has 5 stages. This feature together with the absence of denormals and sticky calculation leads to an area reduction of 25% ($3110\mu m^2$ vs. $2332\mu m^2$). In the first stage, the smallest operand is identified and the difference of exponents is calculated. In the second stage the significand of the smallest operand is shifted and the 2-complement of the smallest operand is calculated if required. The third state carries out the addition of the aligned operands and a possible overflow is detected and corrected. In the forth stage the number of leading zeros are calculated for normalization, which is carried out in the fifth stage. Notice that, unlike the conventional one, a sixth rounding stage is not necessary in HUB since the result of the fifth stage is already normalized and rounded to the nearest HUB number.

In summary, we have integrated a HUB adder in the Sargantana processor, reducing the number of stages and allowing the use of this new format in a RISC-V processor. As future work, we plan to extend the HUB format to all FP RISC-V arithmetic operations.

# References

[1] J. Hormigo and J. Villalba. "New formats for computing with real-numbers under round-to-nearest". In: *IEEE Transactions on Computers* 65.7 (July 2016), pp. 2158–2168. DOI: 10.1109/TC.2015.2479623.

[2] Víctor Soria-Pardos et al. "Sargantana: A 1 GHz+ In-Order RISC-V Processor with SIMD Vector Extensions in 22nm FD-SOI". In: *2022 25th Euromicro Conference on Digital System Design (DSD)*. IEEE. 2022, pp. 254–261.

[3] M. Ercegovac and T. Lang. *Digital Arithmetic*. 1st. Morgan Kaufmann, 2004.